\documentstyle[preprint,aps,prl]{revtex}
\begin{document}

\draft

\title{Anomalous diffusion in the presence of external forces: exact
time-dependent solutions and entropy}

\author{Constantino Tsallis}
\address{Department of Chemistry, Baker Laboratory, and Materials Science
Center\\
Cornell University, Ithaca, NY 14853-1301
\\and\\ Centro Brasileiro de Pesquisas F\'{\i}sicas\\
Rua Xavier Sigaud 150, 22290-180 -- Rio de Janeiro -- RJ, Brazil
\footnote{Permanent address (tsallis@cat.cbpf.br)}}
\author{Dirk Jan Bukman}
\address{Department of Chemistry, Baker Laboratory, Cornell University, Ithaca,
NY 14853-1301}

\date{\today}
\maketitle

\begin{abstract}
The optimization of the usual entropy $S_1[p]=-\int du\; p(u)\; ln\;p(u)$ under
appropriate constraints is closely related to
the Gaussian form of the {\it exact} time-dependent solution of the
Fokker-Planck equation describing an important class of {\it normal}
diffusions.
We show here that the optimization of the generalized entropic form
$S_q[p]=\{1-\int du\;[p(u)]^q\}/(q-1)$ (with $q=1+\mu-\nu \in {\bf \cal{R}}$)
is closely related to the calculation of the {\it exact} time-dependent
solutions of a generalized, nonlinear, Fokker Planck equation, namely
$\frac{\partial}{\partial t}p^\mu= -\frac{\partial}{\partial x}[F(x)p^\mu]
+D \frac{\partial^2}{\partial x^2}p^\nu$, associated with {\it anomalous}
diffusion in the presence of the external force $F(x)=k_1-k_2x$. Consequently,
paradigmatic types of normal ($q=1$) and anomalous ($q \neq 1$) diffusions
occurring in a great variety of physical situations become {\it unified} in a
single picture.
\end{abstract}


Anomalous diffusion is intensively studied nowadays, both theoretically and
experimentally. It is observed,
for instance, in CTAB micelles dissolved in salted water \cite{bouchaud}, the
analysis of heartbeat histograms in a healthy individual \cite{stanley},
chaotic transport in laminar fluid flow of a water-glycerol mixture in a
rapidly rotating annulus \cite{swinney}, subrecoil laser cooling \cite{bardou},
particle chaotic dynamics along the stochastic web associated with a d=3
Hamiltonian flow with hexagonal symmetry in a plane \cite{zaslavsky},
conservative motion in a d=2 periodic potential \cite{klafter}, transport of
fluid in porous media (see \cite{spohn} and references therein), surface growth
\cite{spohn}, and many other interesting physical systems. Its
thermostatistical foundation (as it is known for {\it normal} diffusion) is
naturally highly desirable and has, since long, been looked for (see, for
instance, \cite{montroll} and references therein). This goal was recently
achieved by Alemany and Zanette \cite{alemany} (see also \cite{levy}) for {\it
L\'evy-like} anomalous diffusion, in the context of a generalized, not
necessarily {\it extensive} (additive), thermostatistics that has been
recently proposed \cite{tsallis,curado}. This thermostatistics (already applied
to a considerable variety of physical systems \cite{many} and optimization
techniques \cite{penna}) is based upon the entropic form
\begin{equation}
S_q[p] \equiv \frac{1-\int du\;[p(u)]^q}{q-1}\;\;\;\;\;\;(q \in {\bf \cal{R}})
\end{equation}
which reduces, in the $q \rightarrow 1$ limit, to the standard Boltzmann-Gibbs
entropy
\begin{equation}
S_1[p] \equiv -\int du\;p(u)\;ln\;p(u)
\end{equation}
We show here that an approach similar to that of \cite{alemany}
makes possible a quite satisfactory discussion of a sensibly different
anomalous diffusion, namely of the {\it correlated} type, characterized by the
following generalized,
{\it nonlinear}
Fokker-Planck equation:
\begin{equation}
\frac{\partial}{\partial t}[p(x,t)]^\mu=-\frac{\partial}{\partial x}
\left\{F(x)[p(x,t)]^\mu\right\}+D \frac{\partial^2}{\partial x^2}[p(x,t)]^\nu
\end{equation}
where $(\mu,\nu) \in {\bf {\cal{R}}^2}$, $D > 0$ is a (dimensionless) diffusion
constant, $F(x) \equiv -dV(x)/dx$ is a (dimensionless) external force ({\it
drift}) associated with the potential $V(x)$, and $(x,t)$ is a (dimensionless)
1+1 space-time. Let us mention that, in variance with the correlated type we
are focusing on here, the L\'evy-like anomalous diffusion is associated with a
{\it linear} equation, though in fractional derivatives \cite{zaslavsky}.

We intend to consider here a specific (but very common) drift, namely
characterized by $F(x)=k_1-k_2x$ ($k_1 \in  {\bf {\cal{R}}}$ and $k_2 \geq 0$;
$k_2=0$ corresponds to the important case of constant external force, and
$k_1=0$ corresponds to the so called Uhlenbeck-Ornstein
process\cite{uhlenbeck}).
The particular case $\mu=\nu=1$ corresponds to the standard Fokker-Planck
equation, i.e., to {\it normal} diffusion. The particular case $F(x)=0$ (no
drift) has been considered by Spohn \cite{spohn} for $\mu=1$ and arbitrary
$\nu$ (for instance, $\nu=3$ satisfactorily describes a standard solid-on-solid
model for surface growth), and has been extended by Duxbury\cite{duxbury2} for
arbitrary $\mu$ and $\nu$. The case $(\mu,k_1)=(1,0)$ has been considered by
Plastino and Plastino\cite{plastinos}. Our present discussion recovers {\it
all} of these as particular instances.

First, let us illustrate the procedure we intend to follow, by briefly
reviewing normal diffusion ($\mu=\nu=1$).
We wish to optimize $S_1$ (given by Eq. (2)) with the constraints
\begin{equation}
\int du\;p(u)=1\;\;,
\end{equation}
\begin{equation}
\langle u-u_M \rangle_1 \equiv \int du\;(u-u_M)\;p(u)=0\;\;
\end{equation}
and
\begin{equation}
\langle(u-u_M)^2 \rangle_1 \equiv \int du\;(u-u_M)^2\;p(u) = \sigma^2\;\;,
\end{equation}
$u_M$ and $\sigma$ being fixed {\it finite} real quantities. The optimization
straightforwardly yields the solution
\begin{equation}
p_1(u)=\frac{e^{-\beta(u-u_M)^2}}{Z_1}
\end{equation}
with
\begin{equation}
Z_1 =\int du\;e^{-\beta(u-u_M)^2}=(\pi/\beta)^{1/2}
\end{equation}
where $\beta \equiv 1/T$ is the Lagrange parameter associated with the
constraint (6) and satisfies $\beta=1/(2\sigma^2)$.

On the basis of Eq. (7) we propose, for the $\mu=\nu=1$ particular case of Eq.
(3) (i.e., the standard Fokker-Planck equation), the {\it ansatz}
\begin{equation}
p_1(x,t)=\frac{e^{-\beta(t)[x-x_M(t)]^2}}{Z_1(t)}
\end{equation}
with
\begin{equation}
\frac{\beta(t)}{\beta(0)}=\left[\frac{Z_1(0)}{Z_1(t)}\right]^\lambda
\end{equation}
It follows straightforwardly that $\lambda=2$,
\begin{equation}
\frac{\beta(t)}{\beta(0)}=\left[\frac{Z_1(0)}{Z_1(t)}\right]^2=\frac{1}
{[1-\frac{2D\beta(0)}{k_2}]e^{-2k_2t}+\frac{2D\beta(0)}{k_2}}
\end{equation}
and
\begin{equation}
\frac{dx_M(t)}{dt}= k_1-k_2x_M(t)
\end{equation}
hence
\begin{equation}
x_M(t)=\frac{k_1}{k_2}+\left[x_M(0)-\frac{k_1}{k_2}\right]
e^{-k_2 t}.
\end{equation}
To discuss the $k_2=0$ case we can use $e^{-2k_2t} \sim 1-2k_2t$, which implies
$x_M(t)=x_M(0)+k_1t$ and $1/\beta(t)=[1/\beta(0)]+4Dt$, which, in the limit $t
\rightarrow \infty$ (i.e., $t>>1/[4D\beta(0)]$), yields the familiar result
$1/\beta(t) \sim 4Dt$. This result implies, in turn, the celebrated Einstein
expression $\langle(x-x_M)^2 \rangle_1 \propto t$ for Brownian motion.

Let us now address the {\it general} $(\mu,\nu)$ case. Following along the
lines of
Alemany and Zanette\cite{alemany} (and the generic framework of the generalized
thermostatistics\cite{tsallis,curado}) we now wish to optimize $S_q$ (given by
Eq. (1)). The constraints are Eq. (4),
\begin{equation}
\langle u-u_M \rangle_q \equiv \int du\;(u-u_M)\;[p(u)]^q=0
\end{equation}
(which generalizes Eq. (5)) and
\begin{equation}
\langle(u-u_M)^2 \rangle_q \equiv \int\;(u-u_M)^2\;[p(u)]^q= \sigma^2
\end{equation}
(which generalizes Eq. (6)). This is an appropriate moment for commenting that
the reason for using $[p(u)]^q$ (instead of the familiar $p(u)$) in the
constraints (14) and (15) is the (very essential) fact that by doing so we
{\it preserve}\cite{curado} {\it the Legendre structure of Thermodynamics} and
(through the nonnegativity of $C_q/q$\cite{silva}, where $C_q$ denotes the
specific heat) {\it guarantee thermodynamic stability}. Let us consistently
stress that the constraint (14) is equivalent to
$\langle u \rangle_q= \langle u_M \rangle_q$, but not to $\langle u \rangle_q=
u_M$ (since, unless $q=1$, $\langle u_M \rangle_q \neq u_M$). All these
peculiarities are of course originated by the essential {\it nonextensivity}
that the index $q$ introduces in the theory. For example, if we have two {\it
independent} systems $A$ and $B$ (i.e., $p_{A*B}(u_A,u_B)=p_A(u_A)p_B(u_B)$),
we immediately verify that $S_q(A*B)=S_q(A)+S_q(B)+(1-q)S_q(A)S_q(B)$.

It is straightforward to see that the above described optimization of $S_q$
yields
\begin{equation}
p_q(u)=\frac{[1-\beta(1-q)(u-u_M)^2]^{\frac{1}{1-q}}}{Z_q}
\end{equation}
with
\begin{equation}
Z_q =\int du\;[1-\beta(1-q)(u-u_M)^2]^{\frac{1}{1-q}}
\end{equation}
In the limit $q \rightarrow 1$, these  equations reduce to Eqs. (7) and (8),
respectively. The corresponding {\it ansatz} for solving Eq. (3) now is
\begin{equation}
p_q(x,t)=\frac{\left\{1-\beta(t)(1-q)[x-x_M(t)]^2\right\}^{\frac{1}
{1-q}}}{Z_q(t)}
\end{equation}
with
\begin{equation}
\frac{\beta(t)}{\beta(0)}=\left[\frac{Z_q(0)}{Z_q(t)}\right]^\lambda\;\;\;.
\end{equation}
(as before, $\beta(t)$ and $Z_q(t)$ are nothing but the scaling of space with
time). A tedious (but straightforward) calculation yields
$\lambda= 2\mu$, and $q=1+\mu-\nu$.
An equation for $Z_q(t)$ is also found, namely
\begin{eqnarray}
2\nu D\beta(0)[Z_q(0)]^{2\mu}&-&k_2 [Z_q(t)]^{\mu+\nu} -\nonumber\\
&&\frac{\mu}{\mu+\nu}
\frac{d[Z_q(t)]^{\mu+\nu}}{dt}=0,
\end{eqnarray}
which can be solved by substituting $\widetilde{Z}(t)=Z_q(t)^{\mu+\nu}$.
The resulting solution is ({\it for all values of $k_1$})
\begin{equation}
Z_q(t)=Z_q(0)\left[\left(1-\frac{1}{K_2}\right)e^{-t/\tau}+
\frac{1}{K_2}\right]^
{\frac{1}{\mu+\nu}}
\end{equation}
with
\begin{equation}
K_2 \equiv \frac{k_2}{2\nu D \beta(0)[Z_q(0)]^{\mu-\nu}}
\end{equation}
and
\begin{equation}
\tau \equiv \frac{\mu}{k_2(\mu+\nu)}
\end{equation}
(See Fig. 1). The function $x_M(t)$ is the same as in the case of normal
diffusion,
Eq. (13), since it only describes the motion of the average of the
distribution $p_q(x,t)$, and does not depend on the way in which it spreads.
$\beta(0)$ and $Z_q(0)$ are determined by the initial condition (i.e., by
$p_q(x,0)$).

For $k_2=0$, Eq. (21) becomes
\begin{equation}
Z_q(t)=\left\{[Z_q(0)]^{\mu+\nu}+ \frac{2\nu (\nu+\mu)D\beta(0)
[Z_q(0)]^{2\mu}}{\mu} t
\right\}^{\frac{1}{\mu+\nu}}
\end{equation}
which, for $t \rightarrow \infty$,
asymptotically recovers Duxbury's solution\cite{duxbury2}, namely $1/\beta(t)
\propto [Z_q(t)]^{2\mu} \propto t^{\frac{2\mu}{\mu+\nu}}$. As we see, $\mu/\nu
=1,\;>1$ and $<1$ respectively imply that $[x(t)-x_M(t)]^2$ scales like $t$
({\it normal diffusion}), faster than $t$ ({\it superdiffusion}) and slower
than $t$ ({\it subdiffusion}).  The limits $\mu/\nu=0$ and $\mu/\nu=\infty$
correspond to ``no diffusion'' and ballistic motion, respectively.

For $(\mu,k_1)=(1,0)$, the present set of equations reduces to that of Plastino
and Plastino\cite{plastinos}.

Let us mention that the general solution given in Eq. (21) can be derived from
that for $\mu=1$ (and arbitrary $k_1$)
by defining $\widetilde{p}(x,t)=[p(x,t)]^{\mu}$, and
$\widetilde{\nu}=\nu/\mu$, as can easily be seen from Eq. (3).

Finally, by using Eq. (19) with $\lambda=2\mu$, we can verify that
\begin{equation}
\int dx\;p_q(x,t)= [Z_q(t)/Z_q(0)]^{\mu-1}\;\int dx\;p_q(x,0)\;\;.
\end{equation}
Consequently, the norm ("total mass") is generically conserved for all times
only if $\mu=1$ ($\forall \;K_2$) or if $K_2=1\;(\forall \mu)$. For
$0 \leq K_2<1$ (a common case), the norm monotonically {\it increases
(decreases)} with time if $\mu>1$ ($\mu<1$). If $K_2>1$, it is the other way
around.

Before ending let us mention that, {\it also} when $t$ grows to infinity, the
solutions we have found must be physically meaningful. This imposes
$\mu/\nu>-1$. Indeed, if $k_2 \neq 0$, $\tau$ in Eq. (21) must be positive,
which implies $\mu/\nu>-1$. Also, if $k_2=0$, $x$ must scale with an {\it
increasing} function of $t$; hence, $\beta(t)$ must {\it decrease} with $t$,
which implies (through Eqs. (19) and (24)) $2\mu/(\mu+\nu)>0$, hence, the
already mentioned restriction applies once again. The entire picture which
emerges is indicated in Fig. 2 (we have not focused the $\mu<0$ region because
that would force us to discuss the stability of the solutions with respect to
small departures, and this lies outside of the scope of the present work).

Summarizing, on {\it general} grounds, we have shown that thermostatistics
allowing for nonextensivity constitute a theoretical framework within which a
rather nice {\it unification} of normal and {\it correlated} anomalous
diffusions can be achieved. Both types of diffusions have been founded, on
equal footing, on primary concepts of (appropriately generalized)
Thermodynamics and Information
Theory.

On {\it specific} grounds, we have obtained, for a generic linear force $F(x)$,
the physically relevant {\it exact} (space, time)-dependent solutions of a
considerably generalized Fokker-Planck equation, namely Eq. (3).

It is with pleasure that one of us (C.T.) acknowledges warm hospitality by B.
Widom at the Baker Laboratory.  This work was carried out in the research group
of B. Widom, and was supported by the National Science Foundation and
the Cornell University Materials Science Center.

\begin{figure}
\caption{The $\mu/\nu =1/3$ example:
(a)  Time dependence of $\beta(0)/\beta(t)=[Z_q(t)/Z_q(0)]^{2\mu}$
for $Z_q(0) \neq 0$ and typical values of $K_2$ (indicated at the
right of each curve).  The curve for $K_2 =
0$ lies on the vertical axis.  For $K_2 = 0.25, 0.5$, and $2$ the
asymptotic values for $t/\tau \rightarrow \infty$ are shown by the dashed
lines.  ($\tau$ is defined in Eq. (23).);
(b) Time dependence of $\{\beta(0)[Z_q(0)]^{2\mu}\}/\beta(t)= [Z_q(t)]^{2\mu}$
for $Z_q(0)=0$, $\beta(0)[Z_q(0)]^{2\mu} \neq 0$, and typical values of
$K_2^{'} \equiv k_2/\{2\nu D \beta(0)[Z_q(0)]^{2\mu}\}$
(indicated at the right of each curve).   The curve for $K_2 = \infty$
coincides
with the horizontal axis.  All curves saturate at a finite value as
$t\rightarrow \infty$, except that for $K_2^{'} = 0$, which is
proportional to $t^{2\mu/(\mu+\nu)}$ for all $t$.}
\end{figure}

\begin{figure}
\caption{"Norm conservation" means that $N \equiv \int dx\;p_q(x,t)$ is
time-invariant; "Norm creation" means that $N$ monotonically increases
(decreases) with time if $K_2 <1$ ($K_2>1$); "Norm dissipation" means that $N$
monotonically decreases (increases) with time if
 $K_2 <1$ ($K_2>1$). "Normal diffusion", "Superdiffusion" and "Subdiffusion"
refer to the fact that, for $k_2=0$, $(x-x_M)^2$ scales like $t$, faster than
$t$ and slower than $t$, respectively. The standard Fokker-Planck equation
corresponds to $\mu=\nu=q=1$. For the precise meaning of "unphysical", see the
text. On the $\mu=1$ line we have $q=2-\nu$; consequently, when $\nu$ varies
from $\infty$ to
$-1$, $q$ varies from $-\infty$ to $3$, which precisely is the interval within
which Eq. (4) (and, consistently, $\int dx\;p_q(x,0)=1$) can be satisfied.}
\end{figure}

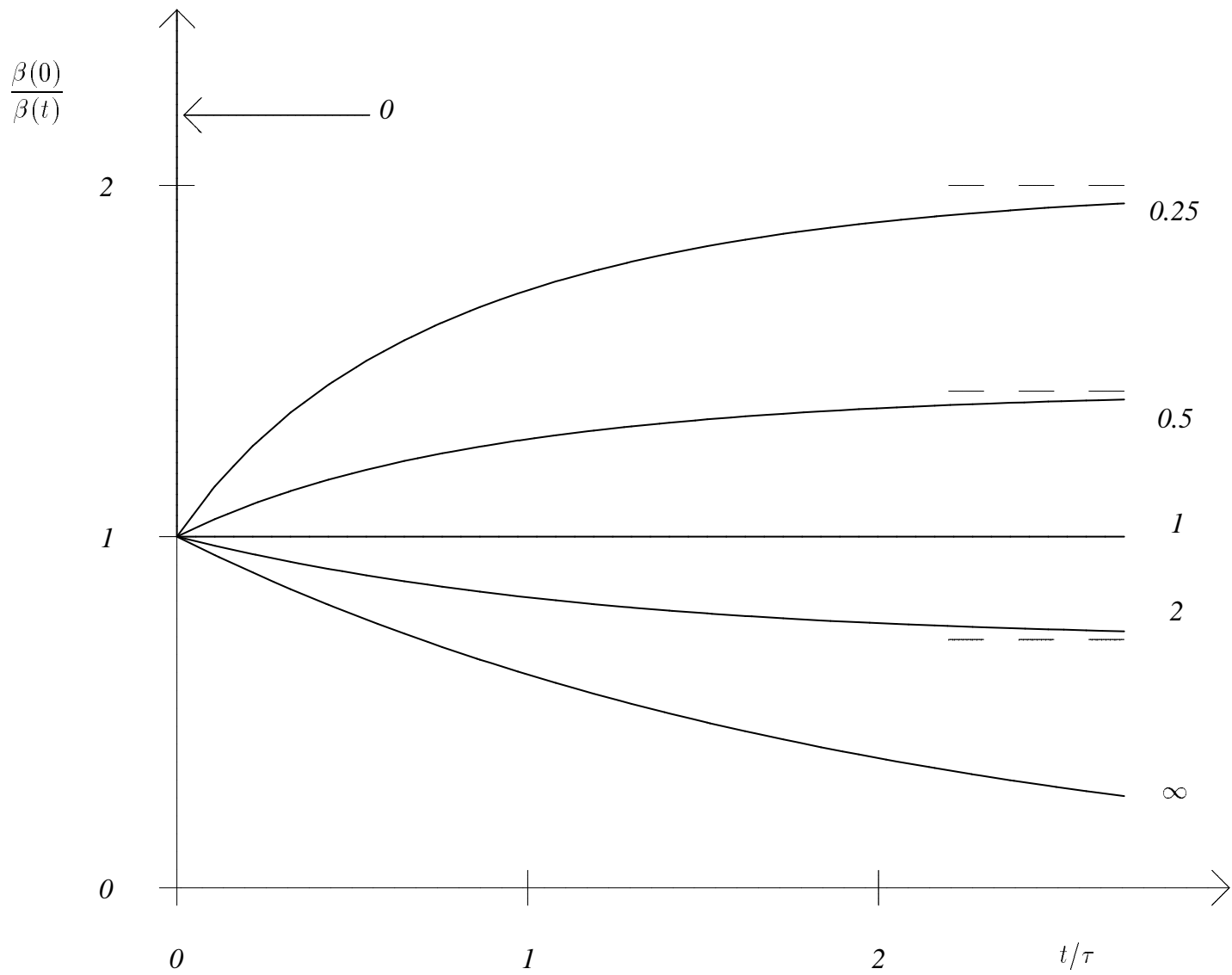
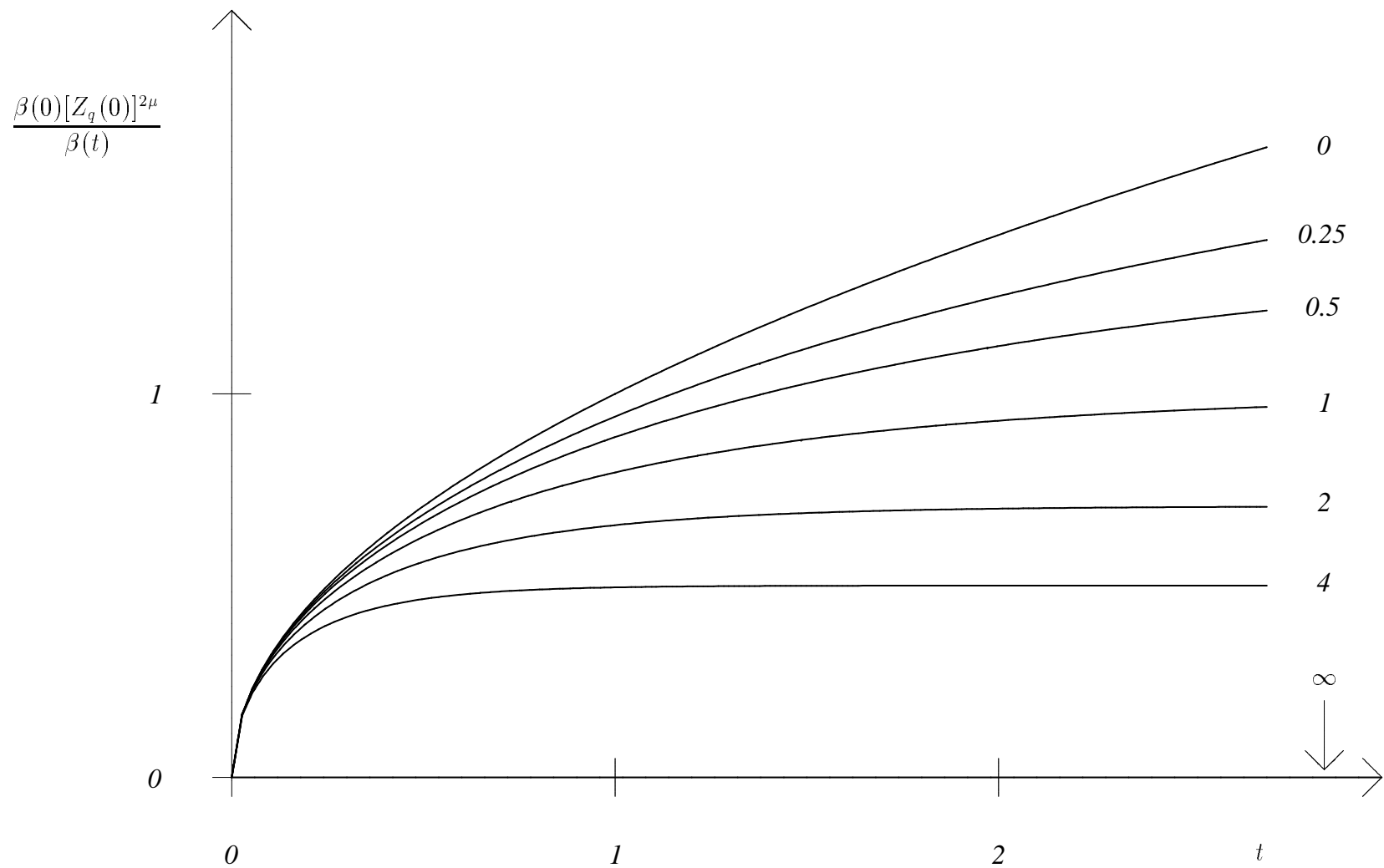
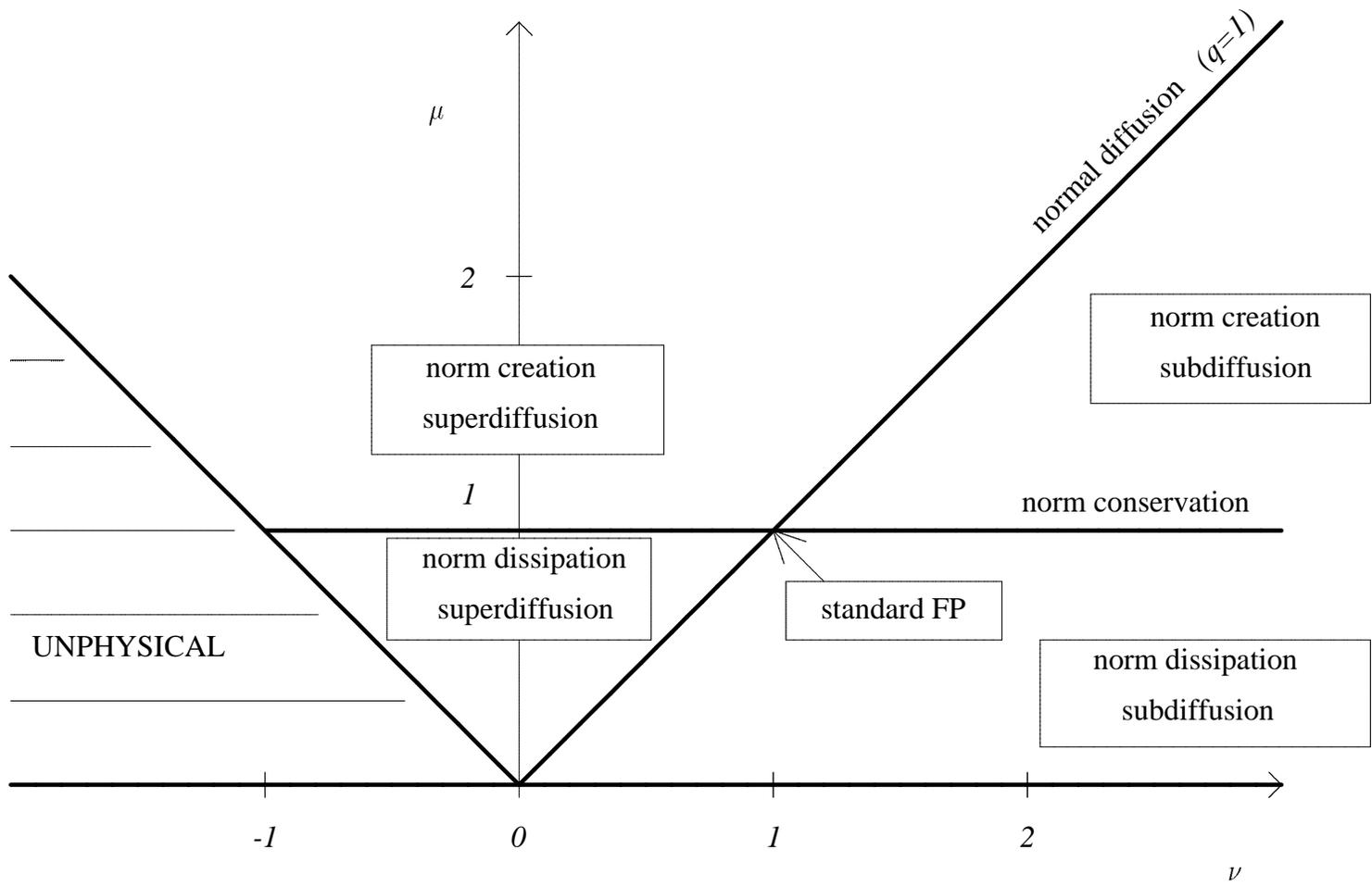
\end{document}